\RequirePackage{fix-cm}
\documentclass[smallextended]{svjour3}       
\smartqed  
\usepackage{graphicx}

\usepackage{amsmath,amssymb}

\newcommand{\e}{{\rm e}}
\newcommand{\ad}{{\rm ad}}

\begin{document}

\title{A note on trigonometric identities involving non-commuting matrices
}


\author{Ana Arnal         \and
        Fernando Casas \and
        Cristina Chiralt 
}


\institute{Ana Arnal \at
              IMAC and Departament de Matem\`atiques \\
              Universitat Jaume I\\
              12071 Castell\'on, Spain\\
              \email{arnal@mat.uji.es}           
           \and
              Fernando Casas \at
              IMAC and Departament de Matem\`atiques \\
              Universitat Jaume I\\
              12071 Castell\'on, Spain\\
              \email{casas@mat.uji.es}           
       \and
              Cristina Chiralt \at
              IMAC and Departament de Matem\`atiques \\
              Universitat Jaume I\\
              12071 Castell\'on, Spain\\
              \email{chiralt@mat.uji.es}           
}

\date{Received: date / Accepted: date}

\maketitle

\begin{abstract}
An algorithm is presented for generating successive approximations to trigonometric functions of sums of non-commuting matrices. The resulting expressions involve nested commutators of the respective matrices. The procedure is shown to converge in the convergent domain of the Zassenhaus formula and can be useful in the perturbative
treatment of quantum mechanical problems, where exponentials of sums of non-commuting skew-Hermitian matrices frequently appear.

\end{abstract}

\section{Introduction}
\label{intro}

Trigonometric matrix functions appear naturally when solving systems of second order
differential equations
\begin{equation}   \label{de.1}
  \frac{d^2 y}{dt^2} + A^2 y = 0, \qquad y(0) = y_0, \qquad y'(0) = y_0',
\end{equation}
whose solution is expressed by
\begin{equation}  \label{de.2}
  y(t) = \cos(t A) y_0 + A^{-1} \sin(t A) y_0'.
\end{equation}
for all $n \times n$ matrices $A$ \cite{higham08fom}. When $A$ is singular, (\ref{de.2}) is interpreted by expanding the matrix
cosine and sine functions as power series of $A$:
\begin{eqnarray}  \label{de.3}
   \cos(A) & = & I - \frac{A^2}{2} + \frac{A^4}{4!} - \frac{A^6}{6!} + \cdots \nonumber \\
   \sin(A) & = & I - \frac{A^3}{3!} + \frac{A^5}{5!} - \frac{A^7}{7!} + \cdots
\end{eqnarray}
Equation (\ref{de.1}) arises in finite element semidiscretizations of the wave equation, whereas similar equations with a non-vanishing
right-hand side of the form $g(t, y(t), y'(t))$ appear in highly oscillatory problems, control  theory, etc.

In this case one has also the matrix analogue of Euler's formula,
\[
   \e^{i A} = \cos(A) + i \sin(A),
\]
so that
\begin{equation}  \label{de.4}
   \cos(A) = \frac{\e^{i A} + \e^{-i A}}{2}, \qquad    \sin(A) = \frac{\e^{i A} - \e^{-i A}}{2i}
\end{equation}
and
\[
   \cos^2(A) + \sin^2(A) = I.
\]
Different algorithms exist in the literature for the practical computation of the matrix cosine and sine (see e.g. \cite{almohy15naf,higham08fom}
and references therein). Several of them make use of the double
angle formula,
\begin{equation}  \label{sq.1}
  \cos(2X) = 2 \cos^2(X) - I,
\end{equation}
to construct an approximation $Y$ to $\cos(A)$ by first considering a matrix $X = 2^{-s} A$ with small norm and then  approximating
$\cos(X)$ by a function $r(X)$ (a truncated Taylor series, a Pad\'e approximant, etc.). $Y$ is then determined by applying formula
 (\ref{sq.1}) $s$ times.

Identity (\ref{sq.1}) is a special case of the addition formulae
\begin{eqnarray}   \label{ad.for1}
  \cos((A+B)t) & = & \cos(At) \, \cos(Bt) - \sin(At) \, \sin(Bt)   \nonumber \\
  \sin((A+B)t) & = & \sin(At) \, \cos(Bt) + \cos(At) \, \sin(Bt)
\end{eqnarray}
which hold if and only if $A B = B A$ \cite[p. 287]{higham08fom}. This is not necessary the case, however, when $t=1$, as the following pair of matrices illustrate
 \cite{frechet52lsn}:
\[
  A = \pi \left( \begin{array}{cr}
    		0  & \  \alpha \\
		-1/\alpha  &  0
	  \end{array}  \right), \qquad\quad
  B = \pi \left( \begin{array}{cc}
    		0  & \  (10 + 4 \sqrt{6}) \alpha \\
		(-10+4 \sqrt{6})/\alpha  &  0
	  \end{array}  \right).	  	
\]
Although $A B \ne B A$ for all $\alpha \ne 0$, a straightforward calculation shows that, indeed, equations (\ref{ad.for1}) with $t=1$ are still valid here.
For general matrices $A$ and $B$, however, one cannot expect them to hold unless their \emph{commutator} $[A,B] \equiv A B - B A$
vanishes. This property is of course related through eq. (\ref{de.4}) with the celebrated Baker--Campbell--Hausdorff (BCH) formula \cite{bonfiglioli12tin}. 
Roughly speaking,  $ \e^A \, \e^B = \e^{A + B + C}$,
where the additional term $C$ is due to the non-commutativity of $A$ and $B$. More in detail, the BCH theorem establishes that $\e^A \, \e^B = \e^Z$, with
\[
  Z = \log(e^A \, \e^B) = A + B  + \sum_{m=2}^{\infty} Z_m(A,B)
\]
and $Z_m(A,B)$ is a linear combination (with rational coefficients) of nested commutators involving $m$ operators $A$ and $B$.  The first terms read explicitly
 \begin{eqnarray*}
 m = 1: \quad   Z_1 & = & A + B \\
 m = 2: \quad   Z_2 & = & \frac{1}{2} [A,B]   \\
 m = 3: \quad  Z_3 & = & \frac{1}{12} [A,[A,B]] - \frac{1}{12}
                 [B,[A,B]]    \\
 m = 4: \quad  Z_4 & = & -\frac{1}{24} [B,[A,[A,B]]]  \\
 m = 5: \quad  Z_5 & = & -\frac{1}{720} [A,[A,[A,[A,B]]]] - \frac{1}{120} [A,[B,[A,[A,B]]]]  \\
   & & - \frac{1}{360} [A,[B,[B,[A,B]]]]  + \frac{1}{360} [B,[A,[A,[A,B]]]] \\
   & & + \frac{1}{120} [B,[B,[A,[A,B]]]] + \frac{1}{720} [B,[B,[B,[A,B]]]].
\end{eqnarray*}
An efficient algorithm for generating explicit expressions of $Z_m(A,B)$ up to an arbitrary $m$ in terms of independent commutators is presented in \cite{casas09aea}.
At this point it is natural to raise the following question: since formulae (\ref{ad.for1}) do not hold in general for non-commutative matrices, is it still possible to express $\cos(A+B)$ in terms of the cosine and sine of $A$ and $B$ for general matrices when $[A,B] \ne 0$? And if the answer is in the affirmative, can this be done
in a systematic (and hopefully efficient) way?

It is the purpose of this note to develop an algorithm that generalizes identities (\ref{ad.for1}) to non-commuting operators, thus providing successive approximations to $\cos(A+B)$
and $\sin(A+B)$ involving $n$-nested commutators of $A$ and $B$ for $n=1,2,\ldots$. As an illustration, if $A$ and $B$ are such that $[A,[A,B]] = [B,[A,B]] = 0$, then
the following exact result holds:
\begin{eqnarray}   \label{ad.for2}
  \cos(A+B) & = & \big( \cos(A) \, \cos(B) - \sin(A) \, \sin(B) \big) \, \e^{\frac{1}{2} [A,B]}   \nonumber \\
  \sin(A+B) & = & \big( \sin(A) \, \cos(B) + \cos(A) \, \sin(B) \big) \, \e^{\frac{1}{2} [A,B]}.
\end{eqnarray}
The algorithm we propose here constitutes in fact a direct application of the dual of the BCH theorem: the so-called Zassenhaus formula,
with multiple applications in quantum mechanical systems and numerical analysis \cite{casas12eco}. 
The problem consists essentially in finding matrices (operators) $C_1, C_2, \ldots$ such that
$\e^{A+B} = \e^A \, \e^B \, \e^{C_1} \, \e^{C_2} \cdots$, with $C_i$ depending only on nested commutators of $A$ and $B$.

Expressions like (\ref{ad.for2}) can be useful in the perturbative treatment of quantum problems where exponentials of sums of non-commuting skew-Hermitian operators
frequently appear \cite{galindo90qme}.

\section{Zassenhaus formula}

To establish the Zassenhaus formula we consider two non commuting indeterminate variables $X$, $Y$ and the free
 Lie algebra generated by them, $\mathcal{L}(X,Y)$. This, roughly speaking, can be viewed as the set of linear combinations of
 all commutators that
 can be constructed with $X$ and $Y$. The elements of $\mathcal{L}(X,Y)$ are called Lie polynomials \cite{postnikov94lga}. A free Lie
 algebra is a universal object, so that results formulated in $\mathcal{L}(X,Y)$ are valid for any (finite- or infinite-dimensional) Lie algebra \cite{munthe-kaas99cia}.

 Let us suppose then that $X, Y \in \mathcal{L}(X,Y)$. The Zassenhaus formula establishes that
 the exponential $\e^{X+Y}$ can be uniquely decomposed as
\begin{equation}  \label{zass.1}
      \e^{X + Y} = \e^X \, \e^Y \, \prod_{n=2}^{\infty} \e^{C_n(X,Y)} = \e^X \, \e^Y \, \e^{C_2(X,Y)} \,
      \e^{C_3(X,Y)} \, \cdots  \, \e^{C_k(X,Y)} \, \cdots,
\end{equation}
where $C_k(X,Y) \in  \mathcal{L}(X,Y)$ is a homogeneous Lie polynomial in $X$ and $Y$ of degree $k$
\cite{magnus54ote,suzuki77otc,weyrauch09ctb,wilcox67eoa,witschel75ooe}. The first terms read explicitly
 \begin{equation}   \label{cs}
\aligned
  &  C_2(X,Y) = -\frac{1}{2} [X,Y] \\
  &  C_3(X,Y) = \frac{1}{3} [Y,[X,Y]] + \frac{1}{6} [X,[X,Y]] \\
  &  C_4(X,Y) = -\frac{1}{24} [X,[X,[X,Y]]] - \frac{1}{8} [Y,[X,[X,Y]]] - \frac{1}{8} [Y,[Y,[X,Y]]].
\endaligned
\end{equation}
A recursive algorithm has been proposed in  \cite{casas12eco} for obtaining the terms
$C_n$ up to a prescribed value of $n$ directly in terms of the minimum number of independent commutators involving $n$ operators $X$ and
$Y$. The procedure, in addition, can be easily implemented in a symbolic
algebra system without any special requirement, beyond the linearity property of the commutator. It reads as follows:
\begin{equation}   \label{alg.1}
\begin{array}{l}
  \mbox{Define} \; f_{1,k} \; \mbox{by}   \\
   \;\;\;  \displaystyle f_{1,k} =  \sum_{j=1}^k \frac{(-1)^k}{j! (k-j)!} \ad_Y^{k-j} \ad_X^j Y, \\
   \displaystyle C_2 = \frac{1}{2}  \, f_{1,1},  \\
   \mbox{Define} \; f_{n,k} \quad n \ge 2, \; k \ge n \;\;  \mbox{by}  \\
   \;\;\; \displaystyle  f_{n,k} = \sum_{j=0}^{[k/n]-1}  \frac{(-1)^j}{j!} \ad_{C_n}^{j} f_{n-1,k-nj}, \\
   \displaystyle C_n = \frac{1}{n} f_{[(n-1)/2],n-1} \quad n \ge 3.
\end{array}
\end{equation}
Here $[k/n]$ denotes the integer part of $k/n$ and the ``ad" operator is defined by
\[
\ad_A B = [A,B], \qquad  \ad_A^j B = [A, \ad_A^{j-1} B], \qquad \ad_A^0 B = B.
\]
Whereas the factorization (\ref{zass.1}) is well defined in the free Lie algebra $\mathcal{L}(X,Y)$, it has only a finite radius of convergence when
$X$ and $Y$ are
 $n \times n$ real or complex matrices. Specifically,
 \begin{equation}   \label{conv.1}
   \lim_{n \rightarrow \infty} \e^{X} \, \e^{Y} \,
   \e^{C_2} \cdots \e^{C_n} = \e^{X+Y}
\end{equation}
only in a certain subset of the plane $(\|X\|, \|Y\|)$ \cite{bayen79otc,suzuki77otc}. As a matter of fact, by bounding appropriately the terms $f_{n,k}$ and
also the $C_n$, i.e., by showing that
\[
   \| f_{n,k} \| \le d_{n,k}, \qquad  \|C_n\| \le \delta_n = \frac{1}{n} d_{[(n-1)/2],n-1}
\]
and analyzing (numerically) the convergence of the series $\sum_{n=2}^{\infty} \delta_n$, it can be shown that the convergence domain    
contains the region
$\|X\|+\|Y\| < 1.054$, and extends to the points $(\|X\|,0)$ and $(0,\|Y\|)$
with arbitrarily large values of $\|X\|$ or $\|Y\|$ \cite{casas12eco}. In practical applications, however, the infinite product  (\ref{zass.1})  is truncated
at some $n$ and then one takes the approximation
 \begin{equation}  \label{zas.app.1}
     \e^{X + Y} \approx \e^X \, \e^Y e^{C_2(X,Y)} \, \e^{C_3(X,Y)} \, \cdots  \, \e^{C_n(X,Y)}.
\end{equation}
When the Zassenhaus formula is applied to $\exp(\pm i (X+Y))$, one gets
\begin{eqnarray}  \label{zass.com.1}
   \e^{i(X + Y)}  & = &  \e^{i X} \, \e^{i Y}  e^{\widehat{C}_2(X,Y)} \, \e^{\widehat{C}_3(X,Y)} \, \e^{\widehat{C}_4(X,Y)} \cdots  \nonumber \\
   \e^{-i(X + Y)}  & = &  \e^{-i X} \, \e^{-i Y}  e^{\widetilde{C}_2(X,Y)} \, \e^{\widetilde{C}_3(X,Y)} \, \e^{\widetilde{C}_4(X,Y)} \cdots,
\end{eqnarray}
respectively, where
\[
   \widehat{C}_n = i^n C_n, \qquad \widetilde{C}_n = (-i)^n C_n, \qquad n \ge 2
\]
and $C_n$ is determined by algorithm  (\ref{alg.1}).   In more detail,
\begin{equation}   \label{zass.com.2}
   \aligned
   &  \widehat{C}_{2k} = \widetilde{C}_{2k}  = (-1)^k \, C_{2k},   \\
   &  \widehat{C}_{2k+1} = -\widetilde{C}_{2k+1} = (-1)^k \, i \, C_{2k+1},  \qquad k \ge 1.
 \endaligned
\end{equation}

\section{The algorithm}

Expansions (\ref{zass.com.1}), together with (\ref{de.4}), allow us to design a recursive procedure and obtain expressions for $\cos(X+Y)$ and $\sin(X+Y)$ in terms of
the sine and cosine of $X$ and $Y$. Since
\[
     \cos(X+Y) = \frac{1}{2} \big( \e^{i (X+Y)} + \e^{-i (X+Y)} \big), \qquad    \sin(X+Y) = \frac{1}{2i} \big( \e^{i (X+Y)} - \e^{-i (X+Y)} \big),
\]
all we have to do is to insert the factorizations (\ref{zass.com.1}) in these expressions and collect terms up to the order $n$ considered. Specifically, let us first introduce
\begin{equation} \label{z1}
 \aligned
  & z_{1,1}  \equiv  \e^{i X} \, \e^{i Y} =  \cos(X) \cos(Y) - \sin(X) \sin(Y)    \nonumber \\
  &  \quad + i \big( \cos(X) \sin(Y) + \sin(X) \cos(Y) \big)  \nonumber \\
  & z_{1,2}  \equiv  \e^{-i X} \, \e^{-i Y} =  z^*_{1,1}
  \endaligned
\end{equation}
and, for $n \ge 2$,
\begin{equation}  \label{zn}
 z_{n,1} = z_{n-1,1} \, \e^{\widehat{C}_n}  \qquad\qquad
z_{n,2} = z_{n-1,2} \, \e^{\widetilde{C}_n}.
\end{equation}
Then it is clear that
\begin{equation}  \label{cszn}
  \aligned
  &  \Psi_n^{[C]}(X,Y) \equiv  \frac{1}{2} ( z_{n,1} + z_{n,2}) \approx \cos(X + Y) \\
  &  \Psi_n^{[S]}(X,Y) \equiv  \frac{1}{2i} ( z_{n,1} - z_{n,2}) \approx \sin(X + Y)
  \endaligned
\end{equation}
Thus, up to $n=2$, one has the approximations
\[
 \aligned
  & \Psi_2^{[C]} =    \frac{1}{2} ( z_{2,1} + z_{2,2}) = \Psi_1^{[C]} \, \e^{-C_2} = \mathrm{Re}(z_{1,1}) \, \e^{-C_2}  \\
  & \Psi_2^{[S]} =    \frac{1}{2i} ( z_{2,1} - z_{2,2}) =  \Psi_1^{[S]} \,  \e^{-C_2} = \mathrm{Im}(z_{1,1}) \, \e^{-C_2}
 \endaligned
\]
which reproduce, with $C_2$ given by (\ref{cs}),  expressions  (\ref{ad.for2}) (with the replacement of $X$, $Y$ by $A$ and $B$, respectively), whereas analogously
\[
  \Psi_3^{[C]} = \Psi_2^{[C]} \, \cos(C_3) + \Psi_2^{[S]} \, \sin(C_3), \qquad \Psi_3^{[S]} = - \Psi_2^{[C]} \, \sin(C_3) + \Psi_2^{[S]} \, \cos(C_3).
\]
The general algorithm can then be established as follows:
\begin{equation}   \label{alg.2}
\begin{array}{l}
  \Psi_1^{[C]}  =  \mathrm{Re}(z_{1,1}) = \cos(X) \cos(Y) - \sin(X) \sin(Y)  \\
  \Psi_1^{[S]} = \mathrm{Im}(z_{1,1}) = \cos(X) \sin(Y) + \sin(X) \cos(Y)  \\
  \mbox{For } k=1,2,\ldots \\
   \;\;\;  \displaystyle \Psi_{2k}^{[C]}  =  \Psi_{2k-1}^{[C]} \, \e^{(-1)^k C_{2k}} \\
   \;\;\;  \displaystyle \Psi_{2k}^{[S]}  =  \Psi_{2k-1}^{[S]} \, \e^{(-1)^k C_{2k}} \\
   \;\;\;  \displaystyle \Psi_{2k+1}^{[C]}  =  \Psi_{2k}^{[C]} \, \cos(C_{2k+1}) - (-1)^k \, \Psi_{2k}^{[S]} \, \sin(C_{2k+1}) \\
   \;\;\;  \displaystyle \Psi_{2k+1}^{[S]}  =  \Psi_{2k}^{[S]} \, \cos(C_{2k+1}) + (-1)^k \, \Psi_{2k}^{[C]} \, \sin(C_{2k+1}).
\end{array}
\end{equation}
Moreover, it is possible to establish the convergence of the procedure as follows. From (\ref{cszn}) we have
\[
   \Psi_n^{[C]}(X,Y) =   \frac{1}{2} \left( \e^{i X} \e^{i Y} \e^{\widehat{C}_2}  \e^{\widehat{C}_3} \cdots  \e^{\widehat{C}_n} + 
        \e^{-i X} \e^{-i Y} \e^{\widetilde{C}_2}  \e^{\widetilde{C}_3} \cdots  \e^{\widetilde{C}_n} \right)
\]
and so
\[
\aligned
  &   \lim_{n\rightarrow \infty} \Psi_n^{[C]}(X,Y) = \frac{1}{2}      \lim_{n\rightarrow \infty}    \e^{i X} \e^{i Y} \e^{\widehat{C}_2} \cdots  \e^{\widehat{C}_n}  + 
  \frac{1}{2}    \lim_{n\rightarrow \infty}   \e^{-i X} \e^{-i Y} \e^{\widetilde{C}_2}   \cdots  \e^{\widetilde{C}_n} \\
  &  = \frac{1}{2} \e^{i (X+Y)} + \frac{1}{2} \e^{-i(X+Y)} =  \cos(X+Y)
\endaligned
\]
in the convergence domain of the Zassenhaus formula (\ref{conv.1}), in particular when $\|X\|+\|Y\| < 1.054$. By applying a similar argument, it is also true that
\[
 \lim_{n\rightarrow \infty} \Psi_n^{[S]}(X,Y) =  \sin(X+Y)
\]
in the same domain.

The recursion (\ref{alg.2}) can be easily programmed with a symbolic algebra package in conjunction with algorithm (\ref{alg.1}) to generate the terms
$C_n$ and thus produce approximations to $\cos(X+Y)$ and $\sin(X+Y)$ up to the desired order $n$. In particular, up to $n=4$ we have
\[
  \aligned
  & \cos(X+Y) \approx  \Big( \big( \cos(X) \, \cos(Y) - \sin(X) \, \sin(Y) \big) \, \e^{-C_2(X,Y)} \, \cos(C_3(X,Y)) +  \\
   &  \qquad \big( \cos(X) \sin(Y) + \sin(X) \cos(Y) \big)  \, \e^{-C_2(X,Y)} \, \sin(C_3(X,Y)) \Big) \, \e^{C_4(X,Y)}  \\
  & \sin (X+Y) \approx  \Big( \big(\sin(X) \, \sin(Y) - \cos(X) \, \cos(Y)   \big) \, \e^{-C_2(X,Y)} \, \sin(C_3(X,Y)) +\\
   &  \qquad \big( \cos(X) \sin(Y) + \sin(X) \cos(Y) \big)  \, \e^{-C_2(X,Y)} \, \cos(C_3(X,Y)) \Big) \, \e^{C_4(X,Y)}
\endaligned
\]

\section{Examples}

Next we collect two particular examples to illustrate the use of, and results obtained by, algorithm (\ref{alg.2}) to approximate $\cos(X + Y)$ and
$\sin(X+Y)$.

\paragraph{Example 1.}

Pauli matrices play an important role in many quantum mechanical problems. They are defined by
\begin{equation}\label{Paulis}
    \sigma_1=\left(
               \begin{array}{ccr}
                 0 & & 1 \\
                 1 & & 0
               \end{array}
             \right), \qquad
     \sigma_2=\left(
                  \begin{array}{ccr}
                    0 & & -i \\
                    i & & 0
                  \end{array}
                \right), \qquad
     \sigma_3=\left(
                \begin{array}{ccr}
                  1 & & 0 \\
                  0 & & -1
                \end{array}
              \right).
\end{equation}
and form a basis of $\mathfrak{su}(2)$, the Lie algebra of $2 \times 2$ skew-Hermitian traceless matrices. They verify
\begin{equation}\label{producpaulis}
   \sigma_{j}\sigma_{k}=\delta_ {jk} I +i\epsilon_{jkl}\sigma_l,
\end{equation}
so that their commutators are given by
\begin{equation} \label{conmusigmas}
  [\sigma_{j},\sigma_{k}]=2i\epsilon_{jkl}\sigma_l,
\end{equation}
where $\epsilon_{jkl}$ denotes the Levi--Civita symbol. It can be shown that
\begin{equation}\label{exppaulis}
  \exp(i\boldsymbol{a}\cdot\boldsymbol{\sigma})=
  \cos ( a ) \,I + i \frac{\sin  (a) }{a} \boldsymbol{a}\cdot\boldsymbol{\sigma},
\end{equation}
where $a = \|\boldsymbol{a}\| = \sqrt{a_1^2 + a_2^2 + a_3^2}$ and $\boldsymbol{\sigma} = (\sigma_1, \sigma_2, \sigma_3)$ \cite{galindo90qme}.

Consider a parameter $\varepsilon > 0$ and let us take $X = \sigma_1$ and $Y = \sigma_3$. Then, direct application of (\ref{exppaulis}) shows that
\begin{equation}\label{cossinalfbet}
  \cos(\varepsilon (X + \beta Y)) = \cos(\varepsilon \lambda) I, \qquad  \sin(\varepsilon( X + \beta Y)) = \frac{\sin(\varepsilon \lambda)}{\lambda} (X + \beta Y)
\end{equation}
with $\lambda = \sqrt{1 + \beta^2}$. On the other hand, algorithm (\ref{alg.2}) applied to this case renders
\begin{equation}\label{cossinpauli}\aligned
\Psi_n^{[C]} & = f_n^{[C]}(\varepsilon,\lambda)I+ g_n^{[C]}(\varepsilon,\lambda)  \, i \,\sigma_2\\
\Psi_n^{[S]} & =  f_n^{[S]}(\varepsilon,\lambda) X+ g_n^{[S]}(\varepsilon,\lambda)  \, Y
\endaligned\end{equation}
with (rather involved) 
explicit expressions for the real functions $f_n^{[C]}$, $f_n^{[S]}$, $g_n^{[C]}$, $g_n^{[S]}$. Notice that a non-vanishing term multiplying $i \, \sigma_2$ appears in the
expression of $\Psi_n^{[C]}$, contrary to the exact solution (\ref{cossinalfbet}). It turns out, however, that $g_n^{[C]}(\varepsilon, \lambda)$ goes to zero when $n \rightarrow \infty$.
Moreover, if a series expansion in powers of $\varepsilon$ of these functions is computed, then we reproduce the exact expressions (\ref{cossinalfbet}) up to the order
considered. Thus, in particular, up to order $\varepsilon^8$ we get
\[
\aligned
  f_8^{[C]}(\varepsilon,\lambda) & = 1- \frac{1}{2} \varepsilon^2 \lambda^2 + \frac{1}{24} \varepsilon^4 \lambda^4 - \frac{1}{720} \varepsilon^6 \lambda^6 + \frac{1}{40320} \varepsilon^8 \lambda^8 
    + \mathcal{O}(\varepsilon^{10}) \\
  g_8^{[C]}(\varepsilon,\lambda) & = \mathcal{O}(\varepsilon^{9}) \\
  f_8^{[S]}(\varepsilon,\lambda) & = \varepsilon - \frac{1}{6} \varepsilon^3 \lambda^2 + \frac{1}{120} \varepsilon^5 \lambda^4 - \frac{1}{5040} \varepsilon^7 \lambda^6 + \mathcal{O}(\varepsilon^9) \\
  g_8^{[S]}(\varepsilon,\lambda) & = \beta \left( \varepsilon - \frac{1}{6} \varepsilon^3 \lambda^2 + \frac{1}{120} \varepsilon^5 \lambda^4 - \frac{1}{5040} \varepsilon^7 \lambda^6  \right) 
    + \mathcal{O}(\varepsilon^9)
 \endaligned
\]

\paragraph{Example 2.}

\begin{figure}[h!]
\centering
  \includegraphics{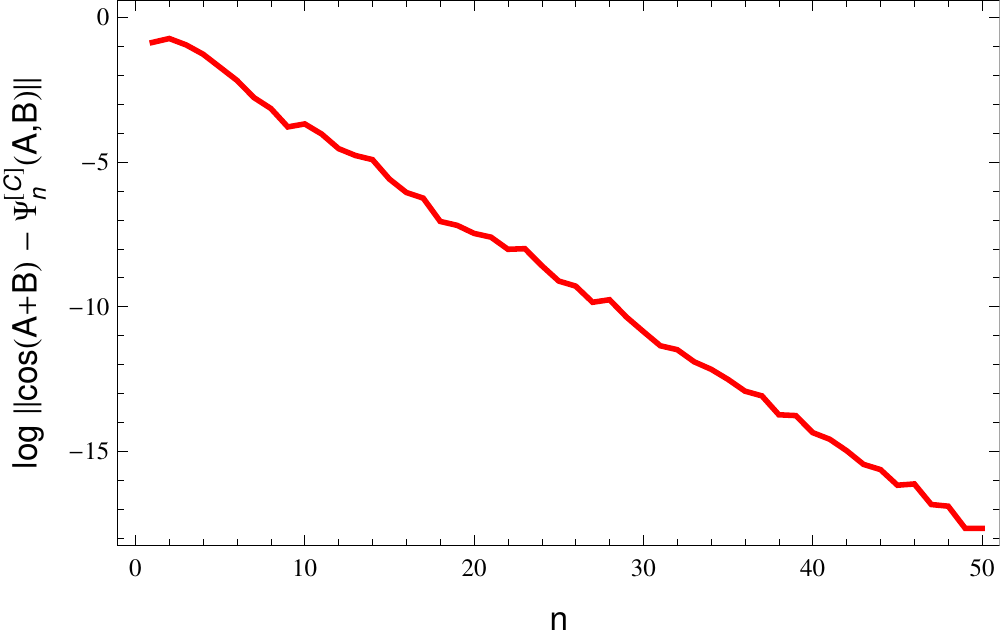}
\caption{Difference between $\cos(A+B)$ and the expansion $\Psi_n^{[C]}(A,B)$ as a function of $n$ for two $10 \times 10$ random matrices $A$ and $B$ with 
$\|A\|_2 = \|B\|_2 = 1$.}
\label{fig.1}       
\end{figure}

For our second example we consider two $10 \times 10$ matrices $A$ and $B$ whose elements are random numbers in the range $(0,1)$ and normalized so that
$\|A\|_2 = \|B\|_2 = 1$. We are therefore \emph{outside} the convergence domain for the Zassenhaus formula guaranteed by \cite{casas12eco}. Then we compute
numerically $\cos(A+B)$ via $X = \e^{i (A+B)}$, $C = \mathrm{Re} \,X$ (with \textit{Mathematica})
 and $\Psi_n^{[C]}(A,B)$ as given by algorithm (\ref{alg.2}) for several values of $n$. Finally we determine
the error  $\log( \| \cos(A+B) - \Psi_n^{[C]}(A,B)\|)$ and represent this value as a function of $n$. In this way we obtain Figure \ref{fig.1}. We clearly observe how the
error decays exponentially with $n$. In other words, algorithm (\ref{alg.2}) provides a convergent expansion for $\cos(A+B)$ well beyond the domain obtained in  \cite{casas12eco}.
A similar conclusion is achieved if one instead considers $\log( \| \sin(A+B) - \Psi_n^{[S]}(A,B)\|)$.

Although algorithm (\ref{alg.2}) is used here to approximate
numerically $\cos(A+B)$, it is by no means intended to be used as a practical alternative to existing numerical procedures to compute the cosine of a matrix, but
rather as an analytical tool in perturbative treatments. This being said, it could also be the case that for certain matrices $A$, $B$, computing the cosine and sine is
 a trivial task, whereas the evaluation of $\cos(A+B)$ and $\sin(A+B)$ is much more involved from a numerical point of view. The idea is then similar to splitting methods
 in the integration of differential equations \cite{blanes16aci}: use $\cos(A)$, $\sin(A)$, $\cos(B)$ and $\sin(B)$ to approximate $\cos(A+B)$ and $\sin(A+B)$.
 In this situation, our procedure could be also competitive with other methods also from the numerical point of view.

\section{Generalizations}

Algorithm (\ref{alg.2}) can be applied of course to get other generalized trigonometric identities involving sums and products of the 
cosine and sine of $X+Y$.  For the
sake of illustration, we next collect the expansions of $\cos (X-Y)-\cos( X+Y)$ and $\sin (X-Y)+\sin( X+Y)$ up to $n=4$ obtained with
our procedure. Specifically,
\[
\aligned
& \cos (X-Y)-\cos( X+Y)   =   \\
 & \qquad  \cos (X) \cos (Y)\, \e^{-C_2(X,-Y) }\cos(C_3(X,-Y)) \,\e^{C_4(X,-Y)} \\
 & \quad -\cos (X) \cos(Y) \, \e^{-C_2(X,Y)} \cos (C_3(X,Y)) \, \e^{C_4(X,Y)}\\
 & \quad -\cos(X) \sin (Y)\,  \e^{-C_2(X,-Y)} \sin(C_3(X,-Y)) \, \e^{C_4(X,-Y)} \\
 & \quad -\cos (X) \sin(Y) \, \e^{-C_2(X,Y)}\sin (C_3(X,Y)) \, \e^{C_4(X,Y)}\\
 & \quad +\sin(X) \cos (Y) \, \e^{-C_2(X,-Y)} \sin(C_3(X,-Y)) \, e^{C_4(X,-Y)} \\
 & \quad -\sin (X)\cos(Y) \, \e^{-C_2(X,Y)} \sin (C_3(X,Y)) \, \e^{C_4(X,Y)}\\
 & \quad +\sin (X) \sin (Y) \, \e^{-C_2(X,-Y)}\cos(C_3(X,-Y)) \, \e^{C_4(X,-Y)} \\
 & \quad +\sin (X) \sin(Y) \, \e^{-C_2(X,Y)} \cos (C_3(X,Y)) \, \e^{C_4(X,Y)}
\endaligned
\]
and 
\[
\aligned
& \sin (X-Y) + \sin( X+Y)   =   \\
 & \quad -\sin (X) \sin (Y) \, \e^{-C_2(X,-Y)}\sin (C_3(X,-Y)) \, \e^{C_4(X,-Y)} \\
 & \quad  +\sin (X)\sin (Y) \, \e^{-C_2(X,Y)} \sin (C_3(X,Y))\, \e^{C_4(X,Y)}\\
 & \quad  -\cos (X) \cos (Y) \, \e^{-C_2(X,-Y)} \sin(C_3(X,-Y)) \, \e^{C_4(X,-Y)} \\
 & \quad  -\cos (X) \cos(Y) \, \e^{-C_2(X,Y)} \sin (C_3(X,Y)) \, \e^{c(4,X,Y)}\\
 & \quad  -\cos (X) \sin (Y) \, \e^{-C_2(X,-Y)} \cos (C_3(X,-Y)) \, \e^{C_4(X,-Y)} \\
 & \quad  +\cos (X) \sin(Y) \, \e^{-C_2(X,Y)} \cos (C_3(X,Y)) \, \e^{C_4(X,Y)}\\
 & \quad  +\sin (X) \cos (Y) \, \e^{-C_2(X,-Y)} \cos(C_3(X,-Y)) \, \e^{C_4(X,-Y)} \\
 & \quad  +\sin (X) \cos(Y) \, \e^{-C_2(X,Y)} \cos (C_3(X,Y)) \, \e^{C_4(X,Y)}.
\endaligned
\]
Notice that if $X$ and $Y$ commute, then $C_n = 0$ for all $n \ge 2$ and  the usual
expressions
\[
\aligned
  &  \cos (X-Y)-\cos( X+Y)   = 2\sin X\, \sin Y, \\
  &  \sin (X-Y) + \sin( X+Y)   = 2\sin X\, \cos Y
\endaligned  
\]   
are recovered.

In the trigonometric expansions obtained with algorithm (\ref{alg.2}) all the successive commutators appear to the right.
This of course is due to the form of the Zassenhaus formula (\ref{zass.1}). There exists, however, an
alternative, ``left-oriented" expression of this formula, namely
\begin{equation}   \label{zass.1.1}
    \e^{X + Y} = \cdots \, \e^{\bar{C}_k(X,Y)} \, \cdots \, \e^{\bar{C}_3(X,Y)} \, \e^{\bar{C}_2(X,Y)} \,
        \e^Y \, \e^X,
\end{equation}
with different but related exponents \cite{casas12eco}:
\[
     \bar{C}_i(X,Y) = (-1)^{i+1} C_i(X,Y),  \qquad i \ge 2. 
\]     
It is then clear that, by using (\ref{zass.1.1}) a similar algorithm can be designed to get alternative expansions for $\cos(X+Y)$ and
$\sin(X+Y)$, this time with commutators appearing to the left. Also invariant expressions with respect to the interchange 
$X \leftrightarrow Y$ can be easily generated by just considering a symmetrized version of the previous expansions. Thus, for instance,
from the first line in eq. (\ref{ad.for2}) we also get
\[
  \cos(A+B) = \cos(B+A) = \big( \cos(B) \, \cos(A) - \sin(B) \, \sin(A) \big) \, \e^{\frac{1}{2} [B,A]} 
\]
and thus
\[
\aligned
   \cos(A+B) & =  \frac{1}{2}  \big( \cos(A) \, \cos(B) - \sin(A) \, \sin(B) \big) \, \e^{\frac{1}{2} [A,B]} \\
    & + \frac{1}{2} \big( \cos(B) \, \cos(A) - \sin(B) \, \sin(A) \big) \, \e^{-\frac{1}{2} [A,B]}.
\endaligned    
\]

\begin{acknowledgements}
This work has been partially supported by Universitat Jaume I trough the project P1-1B2015-16. The second author 
also acknowledges Ministerio de Eco\-no\-m\'{\i}a y Competitividad (Spain) for financial support through 
projects MTM2013-46553-C3 and MTM2016-77660-P (AEI/FEDER, UE).

\end{acknowledgements}

\bibliographystyle{spmpsci}      

\end{document}